\documentclass[aps,prd,preprint,showpacs,showkeys,nofootinbib,amsmath,amssymb]{revtex4-1}
\usepackage{epsf}
\usepackage{color}
\usepackage{dcolumn}
\usepackage{bm}
\everymath{\displaystyle}

\begin{document} 

\title{Zitterbewegung of bosons}

\author{\firstname{Alexander J.}~\surname{Silenko}}
\email{alsilenko@mail.ru} \affiliation{Bogoliubov Laboratory of Theoretical Physics, Joint Institute for Nuclear Research,
Dubna 141980, Russia,\\Institute of Modern Physics, Chinese Academy of
Sciences, Lanzhou 730000, China,\\Research Institute for
Nuclear Problems, Belarusian State University, Minsk 220030, Belarus}

\begin {abstract}
Zitterbewegung of massive and massless scalar bosons and a massive Proca (spin-1) boson is analyzed. The equations describing the evolution of the velocity and position of the scalar boson in the generalized Feshbach-Villars representation
and the corresponding equations for the massive Proca particle in the Sakata-Taketani representation are equivalent to each other and to the well-known equations for the Dirac particle. However, Zitterbewegung does not appear in the Foldy-Wouthuysen representation. Since the position and velocity operators in the Foldy-Wouthuysen representation and their transforms to other representations are the quantum-mechanical counterparts of the corresponding classical variables, Zitterbewegung is not observable.
\end{abstract}


\maketitle

\section{Introduction}

Zitterbewegung belongs to the most important and widely discussed problems of quantum mechanics (QM).
It is a well-known effect consisting in a superfast trembling motion of a free particle. This effect has been first described by Schr\"{o}dinger \cite{ZitterbewegungSc} and is also known for a scalar particle \cite{ZitterbewegungGG,ZitterbewegungFF}. There are plenty of works devoted to Zitterbewegung. In our study, only papers presenting a correct analysis of the observability of this effect are discussed. The correct conclusions
about the origin and observability of this effect have been made in Refs. \cite{ZitterbewegungFF,Zitterbewegungbook,ZitterbewegungOC}. We generalize known results for a scalar boson and carry out a similar analysis for a Proca (spin-1) boson.

The system of units $\hbar=1, c=1$ is used.

\section{Previously obtained results}

The Dirac Hamiltonian for a free particle is given by
\begin{equation}
{\cal H}_{D}=\beta m+\bm\alpha\cdot\bm p
\label{HamD}
\end{equation}
and the Dirac velocity operator has the form
\begin{equation}
\bm v_D\equiv\frac{d\bm r}{dt}=i[{\cal H}_D,\bm r]=\bm\alpha.
\label{Diracvlct}
\end{equation}
We use the standard denotations of the Dirac matrices (see, e.g., Ref. \cite{BLP}).

The operator $\bm v_D$ is time-dependent:
\begin{equation}
\frac{d\bm v_D}{dt}=i[{\cal H}_D,\bm v_D]=i\{\bm\alpha,{\cal H}_D\}-2i\bm\alpha{\cal H}_D=2i(\bm p-\bm\alpha{\cal H}_D).
\label{Diraccel}
\end{equation}

The problem is usually considered in the Heisenberg picture:
\begin{equation}
\bm v_D(t)=e^{i{\cal H}_Dt}\bm\alpha e^{-i{\cal H}_Dt}.
\label{Heispte}
\end{equation}
In the Schr\"{o}dinger picture, the result is the same.
We suppose that 
the eigenvalues of the momentum and Hamiltonian operators are $\bm{\mathfrak{p}}$ and $H$, respectively. In this case, Eq. (\ref{Diraccel}) can be presented in terms of the Dirac velocity operator:
\begin{equation}
\frac{d\bm v_D}{dt}=2i(\bm{\mathfrak{p}}-\bm v_D H).
\label{Dprvelo}
\end{equation}
Its integration shows that the Dirac velocity oscillates:
\begin{equation}
\bm v_D(t)=\left[\bm v_D(0)-\frac{\bm{\mathfrak{p}}}{H}\right]e^{-2iHt}+\frac{\bm{\mathfrak{p}}}{H}.
\label{Dirvele}
\end{equation}
The evolution of the Dirac position operator obtained from this equation is given by
\begin{equation}
\bm r_D(t)=\bm r_D(0)+\frac{\bm{\mathfrak{p}}t}{H}+\frac{i}{2H}\left[\bm v_D(0)-\frac{\bm{\mathfrak{p}}}{H}\right]\left(e^{-2iHt}-1\right).
\label{Dirpoev}
\end{equation}

A similar result has been obtained for a free scalar (spin-0) particle (see Ref. \cite{ZitterbewegungFF} and references therein).
In this case, the initial Feshbach-Villars (FV) Hamiltonian reads \cite{FV}
\begin{equation}
{\cal H}_{FV}=\rho_3m+\left(\rho_3+i\rho_2\right)\frac{\bm p^2}{2m},
\label{HamFV}
\end{equation} where $\rho_i~(i=1,2,3)$ are the Pauli matrices.
The velocity operator in the FV representation is equal to
\begin{equation}
\bm v_{FV}=\left(\rho_3+i\rho_2\right)\frac{\bm p}{m}.
\label{velocFV}
\end{equation}
The corresponding acceleration operator is defined by the equation similar to Eq.
(\ref{Diraccel}) \cite{ZitterbewegungFF}:
\begin{equation}
\frac{d\bm v_{FV}}{dt}=i[{\cal H}_{FV},\bm v_{FV}]=i\{\bm v_{FV},{\cal H}_{FV}\}-2i\bm v_{FV}{\cal H}_{FV}=2i(\bm p-\bm v_{FV}{\cal H}_{FV}).
\label{FVaccel}
\end{equation} It is supposed that 
the eigenvalues of the momentum and Hamiltonian operators are $\bm{\mathfrak{p}}$ and $H$, respectively.
As a result, the final equations of dynamics of the free scalar particle \cite{ZitterbewegungFF} are equivalent to the corresponding equations for the
Dirac particle:
\begin{equation}
\bm v_{FV}(t)=\left[\bm v_{FV}(0)-\frac{\bm{\mathfrak{p}}}{H}\right]e^{-2iHt}+\frac{\bm{\mathfrak{p}}}{H},
\label{FVe}
\end{equation}
\begin{equation}
\bm r_{FV}(t)=\bm r_{FV}(0)+\frac{\bm{\mathfrak{p}}t}{H}+\frac{i}{2H}\left[\bm v_{FV}(0)-\frac{\bm{\mathfrak{p}}}{H}\right]\left(e^{-2iHt}-1\right).
\label{FVpoe}
\end{equation}

However, the presented results have been obtained for the operators defined in the Dirac and FV representations. It has been pointed out in Ref. \cite{OConnell} that the proportionality of the operators $\bm p$ and $\bm v$ can takes place for free particles with any spin. The proportionality of these operators which vanishes the acceleration ($d\bm v/(dt)=0$) can be achieved by the Foldy-Wouthuysen (FW) transformation \cite{FW}. In the FW representation, the Dirac Hamiltonian takes the form \cite{FW}
\begin{equation}
{\cal H}_{FW}=\beta\sqrt{m^2+\bm p^2}, \qquad \bm p\equiv-i\hbar\frac{\partial}{\partial\bm r}
\label{HamFWDp}
\end{equation}
and the velocity operator is given by
\begin{equation}
\bm v_{FW}=\beta\frac{\bm p}{\sqrt{m^2+\bm p^2}}=\frac{\bm p}{{\cal H}_{FW}}.
\label{eqvelmmD}
\end{equation} Similar relations can be obtained for particles with any spin.

It has been shown in Ref. \cite{ZitterbewegungFF} that Zitterbewegung is the result of the interference between positive and negative energy states. It disappears for the ``mean position operator'' \cite{FW} being the position operator in the FW representation \cite{ZitterbewegungFF,Zitterbewegungbook,ZitterbewegungOC}. ``Zitterbewegung was found to be a feature of a particular choice of coordinate
operator associated with Dirac's formulation of relativistic electron theory'' \cite{Zitterbewegungbook}. It can be removed by carrying out the unitary transformation to the FW representation. Experiments do not distinguish between equally valid but different representations leading to the same observables and the transition to the FW representation does not change the physics \cite{ZitterbewegungOC}.
It has been concluded that Zitterbewegung is not an observable \cite{ZitterbewegungOC}. The same conclusion has been made in Refs. \cite{ZitterbewegungDeriglazov,ZitterbewegungKobakhidze}.

Thus, Zitterbewegung is an effect attributed to the Dirac and FV position and velocity operators but not to the corresponding FW operators. However, just the FW position and velocity operators are the quantum-mechanical counterparts of the classical position and velocity (see Refs. \cite{FW,Reply2019,SpinDiracFWR} and references therein). In the Dirac representation, these quantum-mechanical counterparts are defined by the operators \cite{FW}
\begin{equation}
\bm X=\bm r -\frac{\bm\Sigma\times\bm p}{2\epsilon(\epsilon+m)}+\frac{i\bm\gamma}{2\epsilon}-\frac{i(\bm\gamma\cdot\bm p)\bm p}{2\epsilon^2(\epsilon+m)}
\label{meanpos}
\end{equation}
and $d\bm X/(dt)$. For the latter operators and their transforms to other representations, Zitterbewegung does not take place in any representation. The Dirac and FV position and velocity operators are not the quantum-mechanical counterparts of the classical position and velocity (see Refs. \cite{FW,Reply2019,SpinDiracFWR} and references therein). In accordance with 
Refs. \cite{ZitterbewegungOC,ZitterbewegungDeriglazov,ZitterbewegungKobakhidze}, 
Zitterbewegung cannot be observed.

However, there exists an effect which is more or less similar to Zitterbewegung. Feshbach and Villars \cite{FV} have proven that the eigenfunctions of the mean position operator $\bm X$ are not localized in the configuration space but are extended over a radius of the order of the Compton wavelength. These eigenstates are the narrowest possible free wave packets composed only of positive energy states whose behavior agrees with the nonrelativistic (Schr\"{o}dinger) QM. The nonlocality of the particle position takes also place for spinning particles \cite{Zitterbewegungbook,Sakurai}. It has been emphasized by Sakurai \cite{Sakurai} that ``the nonlocality of $\bm X$ is the price we must pay'' for the absence of Zitterbewegung. This indirect connection between the nonlocality of the particle position and Zitterbewegung has also been considered in other works (see, e.g., Refs. \cite{ZitterbewegungRo,ZitterbewegungBD,ZitterbewegungCapelle}). The nonlocality of the particle position manifests in the Darwin (contact) interaction of spinning particles with an electric field.

\section{Zitterbewegung of massive and massless scalar bosons}

In the present study, we generalize the precedent results obtained for a scalar boson.

The transformation fulfilled by Feshbach and Villars \cite{FV} does not applicable for a massless particle. This transformation has been generalized in Ref. \cite{TMP2008}. It has been proposed \cite{TMP2008} to replace the particle mass $m$ in the FV transformation \cite{FV} with the arbitrary nonzero real parameter $N$. For a free massive or massless scalar particle, the wave function of the Klein-Gordon equation, $\psi$, can be presented in the form
\begin{equation}
\psi=\phi+\chi,\qquad i\frac{\partial\psi}{\partial t} =N(\phi-\chi).
\label{KGGFV}
\end{equation}

After multiplying the last relation by $i\partial/(\partial t)$, Eq. (\ref{KGGFV}) can be presented in the matrix form \cite{TMP2008}
\begin{equation}
 i\frac{\partial\Psi_{GFV}}{\partial t} ={\cal H}_{GFV}\Psi_{GFV},\qquad
{\cal H}_{GFV}=\rho_3\frac{\bm p^2+m^2+N^2}{2N}+i\rho_2\frac{\bm p^2+m^2-N^2}{2N}.
\label{HamnGFV}
\end{equation} Here ${\cal H}_{GFV}$ and $\Psi_{GFV}$ are the Hamiltonian and wave function in the generalized Feshbach-Villars (GFV) representation. This representation which has been previously used in Refs. \cite{Honnefscalar,Honnefscalarnew,HonnefscalarLT} is, in fact, an infinite set of representations with different $N$. Nevertheless, the equations of particle dynamics do not depend explicitly on $N$. The velocity operator in the GFV representation is given by
\begin{equation}
\bm v_{GFV}=\left(\rho_3+i\rho_2\right)\frac{\bm p}{N}.
\label{veloGFV}
\end{equation}
The corresponding acceleration operator reads
\begin{equation}
\frac{d\bm v_{GFV}}{dt}=i[{\cal H}_{GFV},\bm v_{GFV}]=i\{\bm v_{GFV},{\cal H}_{GFV}\}-2i\bm v_{GFV}{\cal H}_{GFV}=2i(\bm p-\bm v_{GFV}{\cal H}_{GFV}).
\label{GFVacce}
\end{equation} We can suppose that
the eigenvalues of the momentum and Hamiltonian operators are $\bm{\mathfrak{p}}$ and $H$, respectively. In this case, the dynamics of the scalar particle is defined by the following equations:
\begin{equation}
\bm v_{GFV}(t)=\left[\bm v_{GFV}(0)-\frac{\bm{\mathfrak{p}}}{H}\right]e^{-2iHt}+\frac{\bm{\mathfrak{p}}}{H},
\label{GFVe}
\end{equation}
\begin{equation}
\bm r_{GFV}(t)=\bm r_{GFV}(0)+\frac{\bm{\mathfrak{p}}t}{H}+\frac{i}{2H}\left[\bm v_{GFV}(0)-\frac{\bm{\mathfrak{p}}}{H}\right]\left(e^{-2iHt}-1\right).
\label{GFVpoeq}
\end{equation}
Amazingly, the particle dynamics does not depend on $N$ and is equivalent to that of the
Dirac particle.

The FW transformation for massive and massless ($m=0$) particles results in \cite{TMP2008}
\begin{equation}
{\cal H}_{FW}=\rho_3\sqrt{m^2+\bm p^2}.
\label{HamFW}
\end{equation}
Since the FW velocity operator is proportional to the momentum one,
\begin{equation}
\bm v_{FW}=\rho_3\frac{\bm p}{\sqrt{m^2+\bm p^2}}=\frac{\bm p}{{\cal H}_{FW}},
\label{eqvelmm}
\end{equation}
Zitterbewegung does not appear in the FW representation and is not observable. It does not take place for the FW operators of the position and velocity transformed to other representations. Thus, the more general approach covering a massless scalar particle does not change the conclusions \cite{ZitterbewegungFF,Zitterbewegungbook,ZitterbewegungOC,SpinDiracFWR} about Zitterbewegung.

\section{Zitterbewegung of a Proca boson}

A spin-1 boson can be described by the Proca equations \cite{Pr}
\begin{equation} U_{\mu\nu}=\partial_\mu U_\nu-\partial_\nu U_\mu,\qquad \partial^\nu U_{\mu\nu}-m^2 U_\mu=0.
\label{ProcaII} \end{equation}

The Hamiltonian form of these equations for a free particle is defined by the Sakata-Taketani (ST) transformation \cite{SaTa}. The ST Hamiltonian is given by \cite{SaTa}
\begin{equation}
{\cal H}_{ST}=\rho_3m-i\rho_2\frac{(\bm S\cdot\bm p)^2}{m}+\left(\rho_3+i\rho_2\right)\frac{\bm p^2}{2m}.
\label{HamST}
\end{equation} Here $\rho_i$ are the Pauli matrices and $\bm S=(S_1,S_2,S_3)$ is the spin matrix defined by \cite{YB}
\begin{equation}
S_{1}=\left(\begin{array}{ccc} 0& 0& 0 \\ 0& 0 & -i \\ 0 & i & 0 \end{array}\right), \quad
S_{2}=\left(\begin{array}{ccc} 0& 0& i \\ 0& 0 & 0 \\ -i & 0 & 0 \end{array}\right), \quad
S_{3}=\left(\begin{array}{ccc} 0& -i & 0 \\ i & 0 & 0 \\ 0 & 0 & 0 \end{array}\right).
\label{spinunitmatr}
\end{equation}
This definition is not unique. One can use any other spin matrices satisfying the properties
\begin{equation}
[S_{i},S_{j}]=ie_{ijk}S_{k}, \quad
S_{i}S_{j}S_{k}+S_{k}S_{j}S_{i}=\delta_{ij}S_{k}+\delta_{jk}S_{i},\quad\bm S^2=2{\cal
I},
\label{spinmatrprop}
\end{equation} where ${\cal I}$ is the unit $3\times3$ matrix. The ST transformation cannot be carried out for a massless spin-1 particle.

The ST operators of the velocity and acceleration are given by
\begin{equation}
\bm v_{ST}=\rho_3m+\left(\rho_3+i\rho_2\right)\frac{\bm p^2}{2m},
\label{velocST}
\end{equation}
\begin{equation}
\frac{d\bm v_{ST}}{dt}=i[{\cal H}_{ST},\bm v_{ST}]=i\{\bm v_{ST},{\cal H}_{ST}\}-2i\bm v_{ST}{\cal H}_{ST}.
\label{STaccel}
\end{equation}
Calculations lead to the following result:
\begin{equation}
\{\bm v_{ST},{\cal H}_{ST}\}=2\bm p,\qquad \frac{d\bm v_{ST}}{dt}=2i(\bm p-\bm v_{ST}{\cal H}_{ST}),\qquad
\frac{d\bm v_{ST}}{dt}\Psi_{ST}=2i(\bm{\mathfrak{p}}-\bm v_{ST}H)\Psi_{ST}.
\label{STresul}
\end{equation}

Amazingly, the final equations of dynamics of a free Proca particle are equivalent to the corresponding equations for the Dirac and scalar particles:
\begin{equation}
\bm v_{ST}(t)=\left[\bm v_{ST}(0)-\frac{\bm{\mathfrak{p}}}{H}\right]e^{-2iHt}+\frac{\bm{\mathfrak{p}}}{H},
\label{STe}
\end{equation}
\begin{equation}
\bm r_{ST}(t)=\bm r_{ST}(0)+\frac{\bm{\mathfrak{p}}t}{H}+\frac{i}{2H}\left[\bm v_{ST}(0)-\frac{\bm{\mathfrak{p}}}{H}\right]\left(e^{-2iHt}-1\right).
\label{STpoe}
\end{equation}

For the spin-1 particle, the FW transformation also eliminates Zitterbewegung. The transformed Hamiltonian has the form
\begin{equation}
{\cal H}_{FW}=\rho_3\sqrt{m^2+\bm p^2}
\label{HamFWST}
\end{equation}
and the FW operators of the velocity and momentum become proportional to each other [cf. Eqs. (\ref{eqvelmmD}) and (\ref{eqvelmm})]:
\begin{equation}
\bm v_{FW}=\rho_3\frac{\bm p}{\sqrt{m^2+\bm p^2}}=\frac{\bm p}{{\cal H}_{FW}}.
\label{eqvelST}
\end{equation}
Thus, Zitterbewegung does not appear in the FW representation. As a result, it is unobservable.

\section{Discussion and summary}

We have analyzed Zitterbewegung of the massive and massless scalar bosons and the massive Proca (spin-1) boson. The equations (\ref{GFVe}) and (\ref{GFVpoeq}) describing the evolution of the velocity and position of the scalar boson in the GFV representation
and the corresponding equations (\ref{STe}) and (\ref{STpoe}) for the massive Proca particle in the ST representation are equivalent to each other and to the well-known equations (\ref{Dirvele}) and (\ref{Dirpoev}) for the Dirac particle. In connection with this equivalence, we can mention
the existence of bosonic symmetries of the standard Dirac equation \cite{Simulik,Simulik1,Simulik2,Simulik3,Simulik4,Simulik5}. However, Zitterbewegung does not appear in the FW representation. The position and velocity operators in the FW representation and their transforms to other representations are the quantum-mechanical counterparts of the corresponding classical variables. Therefore, Zitterbewegung takes place \emph{only} for operators which are not the quantum-mechanical counterparts of the classical position and velocity. As a result, Zitterbewegung is not observable. This conclusion agrees with the conclusions about Zitterbewegung previously made in Refs. 
\cite{ZitterbewegungFF,Zitterbewegungbook,ZitterbewegungOC,ZitterbewegungDeriglazov,ZitterbewegungKobakhidze,SpinDiracFWR}.

It has been proven \cite{FV} that the eigenfunctions of the mean position operator $\bm X$ are not localized in the configuration space but are extended over a radius of the order of the Compton wavelength. This effect is more or less similar to Zitterbewegung. The nonlocality of the particle position takes also place for spinning particles \cite{Zitterbewegungbook,Sakurai}. This nonlocality is indirectly connected with Zitterbewegung (see, e.g., Refs. \cite{Sakurai,ZitterbewegungRo,ZitterbewegungBD,ZitterbewegungCapelle}) and manifests in the Darwin (contact) interaction of spinning particles with an electric field.

\begin{acknowledgments}
This work was supported in part by the Belarusian Republican Foundation
for Fundamental Research
(Grant No. $\Phi$18D-002), by the National Natural Science
Foundation of China (Grants No. 11575254 and No. 11805242), and
by the National Key Research and Development Program of China
(No. 2016YFE0130800).
%
The author also acknowledges hospitality and support by the
Institute of Modern
Physics of the Chinese Academy of Sciences. 
\end{acknowledgments}


\end{document}